# Generation and quantum control of giant plasmon pulses by transient quantum coherence

Weiguang Huo,[1, 2] Dmitri V. Voronine[2, 3*] and Marlan Scully[2,3,4]

[1] *Xi'an Jiaotong University, Xi'an, Shaanxi 710049, China*

[2] *Texas A&M University, College Station, TX 77843, USA*

[3] *Baylor University, Waco, TX 76798, USA*

[4] *Princeton University, Princeton, NJ USA*

Amplified ultrashort laser pulses are useful in many fields of science and engineering. Pushing the frontiers of ultrashort pulse generation will lead to new applications in biomedical imaging, communications and sensing. We propose a new, quantum approach to ultrashort pulse generation using transient quantum coherence which predicts order of magnitude stronger pulses generated with lower input energy than in the steady-state regime, reducing the practical heating limitations. This femtosecond quantum-coherent analog of nanosecond Q-switching is not limited by the pulse duration constraints of the latter, and, in principle, may be used for a variety of lasers including x-ray and plasmon nanolasers. We apply this approach to generation of giant plasmon pulses and achieve quantum control of plasmon relaxation dynamics by varying the drive pulse delay, amplitude and duration. We provide insights into the control mechanisms, and discuss future implementations and applications of this new source of ultrashort nanooptical fields.

## I.  INTRODUCTION

Ultrashort laser pulses are the fastest events that have been measured[1,2]. Various methods exist to generate laser pulses in a broad range of pulse durations from nanosecond to attosecond, and frequencies from infrared to x-rays[3-5]. Active research continues improving the performance of ultrafast lasers with new materials, higher powers, shorter durations, smaller sizes and controllable pulse shapes[6]. Large peak amplitudes of ultrashort laser pulses enable many exciting nonlinear optical applications such as high harmonics generation5 , biomedical imaging[7] and laser surgery[8], optical nanofabrication[9], coherent multidimensional optical spectroscopy[10], quantum-coherent nanoscale sensing[11,12], and others. Many new applications are envisioned with the improved performance of shorter, brighter and more energetic pulses such as ultrafast imaging of electron dynamics in energy and charge transfer devices and photosynthetic complexes, of protein folding dynamics, sub-nanometer lithography, and ultrafast classical and quantum information. Various challenges, however, need to be addressed such as improving material properties and pulse generation methods to extend the current ultrashort laser pulse generation technology to wider ranges of parameters and overcome the performance limits.

Here we propose a new, quantum approach to ultrashort pulse generation based on transient quantum coherence. Lasers can be described as quantum heat engines (QHE) in which incoherent thermal radiation is converted into useful work in the form of coherent directional radiation[13]. The QHE power can be increased by laser-driven or noise-induced quantum coherence[14,15]. The quantum coherence can break detailed balance and modify energy level populations of the laser gain medium. Lasers without inversion (LWI) operate based on similar principles[16-19]. Transient LWI has also been investigated[20]. We propose to use transient quantum coherence induced by a laser drive pulse to break the detailed balance on an ultrashort time scale and modify the Q-factor, thereby, generating amplified ultrashort pulses. Our approach is analogous to generation of giant laser pulses by conventional Q-switching[21-23] but it is not bounded by the temporal limitations of the latter which relies on the limitations of electronics, material properties of saturable absorbers, and other factors preventing generation of ultrashort sub-picosecond pulses.

We apply this quantum-coherent Q-switching to ultrashort plasmon pulse generation in subwavelength metal nanostructures. Amplified plasmon pulses may be useful in ultrafast nano-optics[24,25] where resonant optical nanoantennas[26,27] amplify weak ultrafast optical signals from single molecules[28,29] and plasmonic nano-circuits[30,31]. Nanoscale coherent light sources such as nanolasers and surface plasmon generators/amplifiers (spasers) have been proposed[32-35] and recently experimentally realized as proof-of-principle demonstrations[36-40]. However, the pulse duration in these devices was long, and their practical use is limited due to small efficiency and heating/melting effects[41]. To improve the performance we recently proposed to use quantum coherence in three-level gain media coupled to a silver nanoparticle and predicted an order of magnitude increase of the steady-state number of plasmons[42]. Strong continuous drive was used to generate the coherence which could also limit certain applications which are sensitive to heating. Here, we predict new transient quantum coherence effects induced by an ultrashort drive pulse. We achieve two orders of magnitude enhancement of the plasmon peak amplitude compared to the case without a drive. Furthermore, we vary the drive pulse delay, amplitude and duration, and achieve quantum control of the ultrafast plasmon dynamics.

## II. RESULTS

**Generation of giant plasmon pulses by quantum-coherent Q-switching**

The quantum-coherent plasmon pulse generation process is shown in Fig.1. A plasmonic metallic nanostructure such as a silver nanosphere is covered by a layer of the gain medium (Fig.1a) made of three-level quantum emitters such as atoms, molecules or quantum dots, which have a ground state $|1\rangle$ and two excited states $|2\rangle$ and $|3\rangle$ (Fig.1b). States $|1\rangle$ and $|3\rangle$ are coupled by an incoherent pump. An external coherent source of short laser pulses, drives the transition $|2\rangle \rightarrow |3\rangle$. The transition $|2\rangle \rightarrow |1\rangle$ occurs spontaneously and is nearly resonant with the plasmon mode of the nanosphere, and is used to transfer energy from the gain medium to plasmons. As a result, a giant plasmon pulse is generated (Fig.1c).

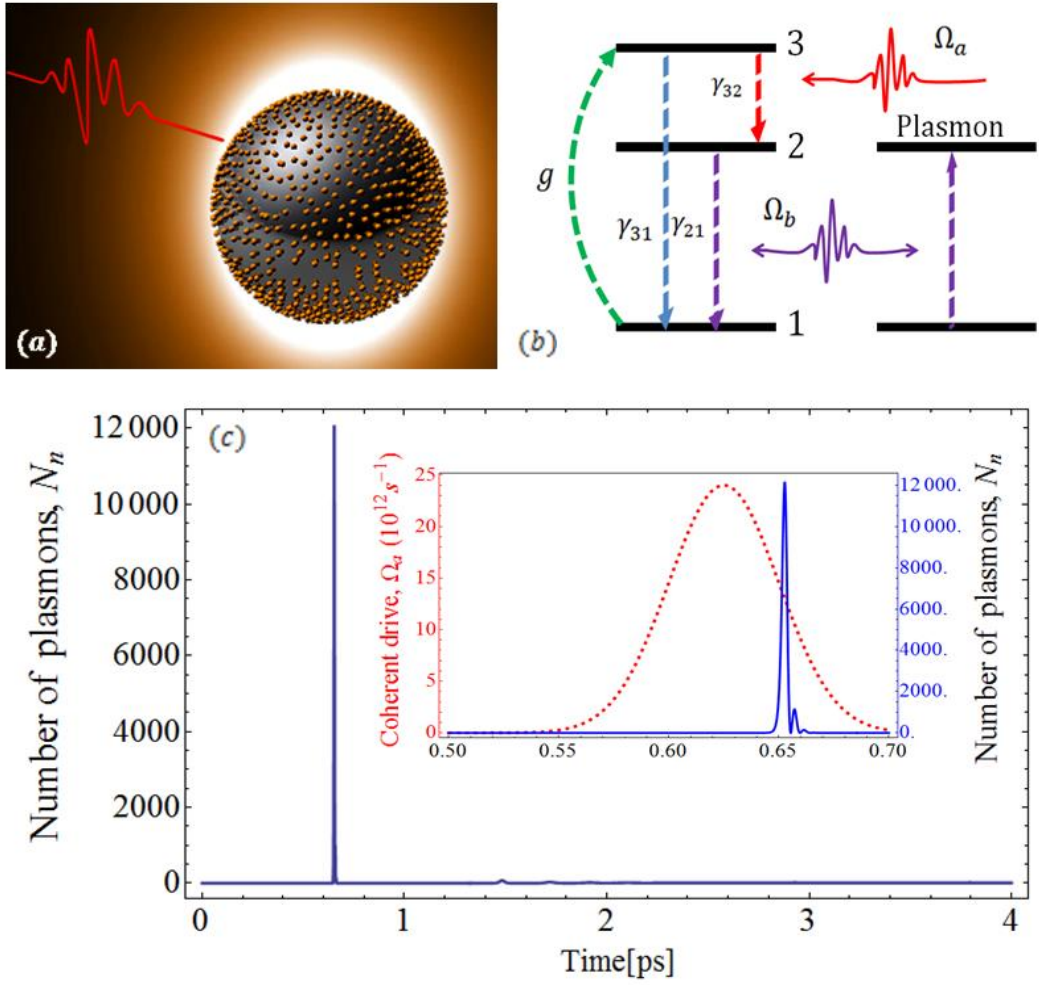

Figure 1. Generation of giant plasmon pulses by quantum-coherent Q-switching. (a) A silver nanosphere surrounded by three-level quantum emitters placed in the near field of a dipolar plasmon mode driven by an ultrashort laser pulse. (b) Energy level diagram reveals the physical mechanisms involved in the quantum-coherent plasmon pulse generation process. (c) Temporal profile of the number of plasmons (solid blue) with a drive pulse (dashed red) arriving at the $\tau_d$=0.625 ps delay time after switching of the pump. The inset shows the zoom-in and reveals that the generated giant plasmon pulse is significantly shorter than the drive pulse.

Using semiclassical theory, we treat the gain medium quantum mechanically and the plasmons and photons classically[42]. We consider the plasmon and photon fields $a_n = a_{0_n} e^{-i\nu_a t}$ and $b_m = b_{0_m} e^{-i\nu_b t}$, respectively, where $a_{0_n}$ and $b_{0_m}$ are slowly varying amplitudes. The

Hamiltonian can be written as

$$\mathcal{H}_{int} = \sum_p \{-\hbar\Delta_b^{(p)}|1\rangle\langle 1| + \hbar\Delta_a^{(p)}|3\rangle\langle 3| - (\hbar\Omega_b^{(p)}|2\rangle\langle 1| + \hbar\Omega_a^{(p)}|3\rangle\langle 2| + c.c)\}, \quad (2)$$

where $\Omega_b^{(p)} = -A_n \boldsymbol{d}_{21}^{(p)} \nabla \phi_n(\mathbf{r}_p) a_{0_n}/\hbar$ is the Rabi frequency for the spasing transition $|2\rangle \to |1\rangle$, and $\Omega_a^{(p)} = -\mathbf{E}_m(\mathbf{r}_p) \boldsymbol{d}_{23}^{(p)} b_{0_m}/\hbar$ is Rabi frequency for the driving transition $|2\rangle \to |3\rangle$. The summation is over all the $p$th chromophores. $\Delta_a$ and $\Delta_b$ are detunings, defined as $\Delta_a = \omega_{32} - \nu_a$ and $\Delta_b = \omega_{21} - \nu_b$. We assume the Gaussian amplitude of the drive pulse laser $\Omega_a^{(p)}$,

$$\Omega_a^{(p)}(t) = \Omega_{a0}^{(p)} e^{-\frac{(t-\tau_d)^2}{2\sigma_d^2}}, \quad (3)$$

and that the driving field is strong enough that $\Omega_{a_0}^{(p)}$ is constant; $\tau_d$ is the drive pulse delay, with respect to the switching time of the pump, and $\sigma_d$ is the drive pulse duration.

The density matrix elements $\rho_{ij}$ satisfy the Liouville-von Neumann equation

$$\dot{\rho}^{(p)} = -\frac{i}{\hbar}[\mathcal{H}_{int}, \rho^{(p)}] - \mathcal{L}\rho^{(p)}, \quad (4)$$

where $\mathcal{L}$ is the dissipative superoperator. The full set of coupled differential equations is given by

$$\dot{\rho}_{11} = \gamma_{21}\rho_{22} + \gamma_{31}\rho_{33} - g\rho_{11} + i(\Omega_b^*\rho_{21} - \Omega_b\rho_{21}^*),$$
$$\dot{\rho}_{33} = -(\gamma_{31} + \gamma_{32})\rho_{33} + g\rho_{11} - i(\Omega_a^*(t)\rho_{32} - \Omega_a(t)\rho_{32}^*),$$
$$\dot{\rho}_{21} = -\Gamma_{21}\rho_{21} - i\Omega_b(\rho_{22} - \rho_{11}) + i\Omega_a^*(t)\rho_{31},$$
$$\dot{\rho}_{31} = -\Gamma_{31}\rho_{31} - i\Omega_b\rho_{32} + i\Omega_a(t)\rho_{21},$$
$$\dot{\rho}_{32} = -\Gamma_{32}\rho_{32} - i\Omega_a(t)(\rho_{33} - \rho_{22}) - i\Omega_b^*\rho_{31},$$
$$\rho_{11} + \rho_{22} + \rho_{33} = 1,$$

where the relaxation rates $\Gamma_{21} = \frac{1}{2}(\gamma_{21} + g) + \gamma_{ph} + i\Delta_b$, $\Gamma_{31} = \frac{1}{2}(\gamma_{21} + \gamma_{31} + g) + \gamma_{ph} + i(\Delta_a + \Delta_b)$, and $\Gamma_{32} = \frac{1}{2}(\gamma_{31} + \gamma_{21} + \gamma_{32}) + \gamma_{ph} + i\Delta_a$. Here $\gamma_{ij}$ are the decay rates for populations, $\gamma_{ph}$ is the phase relaxation rate of the coherence $\rho_{ij}$, and g is the incoherent pump rate.

The corresponding time evolution equation for the plasmonic field is obtained using the Heisenberg equation of motion

$$\dot{a}_{0_n} = -\Gamma_n a_{0_n} + i\sum_p \rho_{21}^{(p)} \widetilde{\Omega}_b^{(p)},$$

where $\Gamma_n = \gamma_n + i\Delta_n$, $\widetilde{\Omega}_b^{(p)} = \Omega_b^{(p)}/a_{0_n}$ is a single pasmon Rabi frequency. We assume that the Rabi frequencies are the same for all chromophores and omit the index (p) below.

In the numerical simulations we used the following parameters[42]. The detuning of the gain medium from the plasmon mode is $\hbar(\omega_{21} - \omega_n)$=0.002 eV. The external dielectric has the permittivity of $\epsilon_d$=2.25. The nanosphere plasmon damping rate $\gamma_n$=5.3×10$^{12}$ s$^{-1}$ with $\Delta_n$=3×10$^{12}$ s$^{-1}$. The gain medium quantum emitters have the following decay rates: $\gamma_{21}$=4×10$^{12}$ s$^{-1}$, $\gamma_{31}$=4×10$^{10}$ s$^{-1}$, $\gamma_{32}$=4×10$^{11}$ s$^{-1}$; the detunings $\Delta_a$ and $\Delta_b$, and the dephasing rate $\gamma_{ph}$ were set to zero. In all simulations we used the incoherent pump rate, g=8×10$^{12}$ s$^{-1}$. The seed plasmon field at time zero was set to 10$^{-5}$.

As an example, we consider a silver nanoparticle of 40 nm radius and the corresponding surface plasmon resonance (SPR) at $\hbar\omega_n$=2.4 eV[42]. The system parameters (energies and relaxation rates) are given in the Methods section. The results of numerical simulations are shown in Fig. 1c. The solid blue curve shows the temporal profile of the generated number of plasmons from a silver nanosphere driven by a short coherent laser pulse at the $|2\rangle \rightarrow |3\rangle$ transition far-detuned from the SPR, with a duration of $\sigma_d$=25 fs and the peak Rabi frequency $\Omega_{a0}$=24×10$^{12}$ s$^{-1}$ (dashed red) arriving at the $\tau_d$=0.625 ps delay time after switching of the pump. The generated giant pulse has a number of plasmons reaching 12000 which is an order of magnitude larger than using a cw drive with the same $\Omega_{a0}$ and two orders of magnitude larger than without drive[42].

The inset of Fig. 1c shows the profile of the Gaussian drive pulse (dashed red) and of the plasmon pulse (solid blue). Interestingly, the plasmon pulse duration (~2 fs) is significantly shorter than the drive pulse duration (25 fs). This shows the possibility of the new approach to generate ultrashort plasmon pulses using transient quantum coherence. The inset of Fig.1c also shows a time delay of the peak of the plasmon pulse with respect to the peak of the drive pulse, $\tau_0 \approx$ 28 fs. We set this delay to be positive if the first plasmon peak is emitted after the peak of the drive pulse. One can also see smaller pulses in Fig.1c and inset. They are due to complex population relaxation dynamics. Our goal was to suppress these small pulses and to deposit more energy of the drive pulse into a single giant plasmon pulse. In the following, we show that the parameters of the drive pulse may be used to control and optimize the properties of the generated plasmons. We investigate the dependence of the plasmon peak intensity on the drive pulse delay, amplitude and duration and discuss the underlying mechanisms of the quantum control.

First, we consider the case without the drive pulse. The simulation results are shown in Fig. 2.

Fig. 2a shows the temporal population evolutions of levels $|1\rangle$, $|2\rangle$ and $|3\rangle$. Initially all the population is in the ground state $|1\rangle$, which then is transferred to $|3\rangle$ by the pump, and subsequently decays to $|2\rangle$ and $|1\rangle$. This leads to the population inversion on the $|2\rangle \rightarrow |1\rangle$ transition, $n_{21}$, shown in Fig. 2b (dashed red) and surface plasmon emission (solid blue) with a number of plasmons $N_n$ in the mode n. Before the system can reach equilibrium it undergoes a sequence of population relaxations leading to a train of short pulses. The physics of these population relaxations is the same as in the conventional lasers[21]. Fig.2b shows that when the population inversion reaches a threshold value the system emits a plasmon pulse and the population inversion drops down. When the plasmon field reaches the level below the steady-state equilibrium the population inversion starts increasing again leading to the generation of the second plasmon pulse, and so on. The difference between this amplified plasmon system (spaser) and the conventional lasers are the dynamical time scales which are determined by the transition rates. Relaxation rates of surface plasmons are on the femtosecond time scale. The gain medium relaxation rates are also fast due to the coupling to plasmons via the Purcell effect. This leads to faster dynamics and faster relaxation oscillations, and a possibility to achieve ultrashort pulses and femtosecond Q-switching. The system dynamics can be alternatively visualized in the phase plane which shows circular traces in the $n_{21}$ - $N_n$ plot converging to a stable equilibrium point (Fig. 2c). The phase-plane description provides a useful clear view of the relation between the population inversion and number of plasmons.

Comparison of Figs. 1c and 2b shows more than two orders of magnitude increase of the plasmon pulse amplitude due to the short drive pulse. This effect is caused by the ultrafast suppression of losses in the gain medium by the transient quantum coherence and by the induced population transfer from level $|3\rangle$ to $|2\rangle$. The number of plasmons depends on the polarization of the gain medium which is related to the coherence $\rho_{21}$:

$$\dot\rho_{21} = -\Gamma_{21}\rho_{21} - i\Omega_b(\rho_{22} - \rho_{11}) + i\Omega_a^*(t)\rho_{31}. \qquad (1)$$

This equation and the parameters are described in the Method section. The number of plasmons can be controlled therefore by two mechanisms: directly by the drive field $\Omega_a$ (third term of Eq. (1)) and via the population inversion (second term of Eq. (1)) which also depends on $\Omega_a$ (see the Methods section). The large population inversion increases the imaginary part of $\rho_{21}$, which decreases the plasmon field amplitude and leads to plasmon relaxation oscillations. We denote this as the *population-inversion* control mechanism. The third term of Eq. (1) shows that the number of plasmons can be directly controlled by the drive $\Omega_a$ in the absence of population inversion. We denote this as the *LWI* control mechanism. These two control mechanisms provide an excellent understanding of the results. The competition between these two mechanisms may be controlled by varying the properties of the drive pulse such as the time delay, amplitude and duration, leading to the effects similar to Q-switching used to generate giant nanosecond pulses in lasers. The conventional Q-switching is achieved via various methods using, for example, acoustooptical, electrooptical or saturable absorber loss modulation. Here we use transient quantum coherence induced by a short drive laser pulse to achieve femtosecond Q-switching. Next we vary the properties of the drive pulse to optimize the plasmon pulse generation.

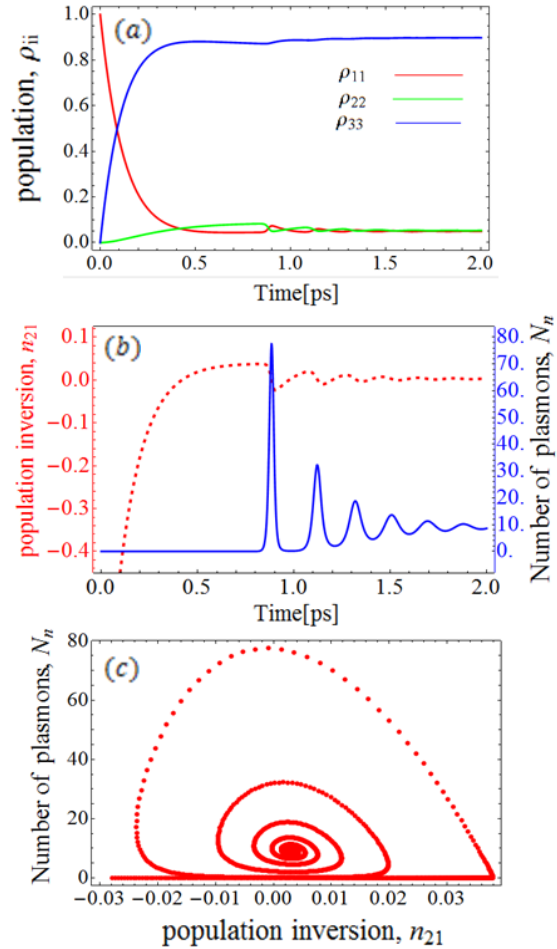

Figure 2. Plasmon relaxation oscillations without a drive. (a) Temporal evolutions of the gain medium level populations. (b) Temporal profile of the population inversion on the $|2\rangle \to |1\rangle$ transition, $n_{21}$ (dashed red) and the number of generated plasmons, $N_n$ (solid blue). (c) Phase plane description of the relation between population inversion and number of plasmons at various points in time.

**Control by the drive pulse delay**

First we perform the quantum control of plasmon pulse generation by varying the time delay of the drive laser pulse with respect of switching of the pump. We show the peak intensity of the plasmon pulse, (i.e. the maximum number of plasmons) at different time delays of the drive pulse with duration $\sigma_d$=25 fs and peak Rabi frequency $\Omega_{a0}$=24×10$^{12}$ s$^{-1}$ in Fig. 3a. The maximum number of plasmons with the drive (solid red) is compared with the number of plasmons without the drive (solid blue). The maximum number of plasmons is anti-correlated with the number of plasmons without the drive. We find three different regimes: (I) the *first* regime showing the largest plasmon pulse amplitude; (II) the second, *spiking* regime with oscillating maximum number of plasmons; and (III) the third, *stable* regime with a constant maximum number of plamons. These three regimes are separated by vertical dashed black lines in Fig. 3a.

The anti-correlation behavior can be explained using the population evolutions, $\rho_{ii}$, population inversion, $n_{21}$, plasmon pulse profiles, $N_n$, and phase plane plots for different time delays for the first, $\tau_d$=0.4 ps, for the spiking $\tau_d$=0.9 ps, and for the stable regime $\tau_d$=3 ps shown in Figs. 3b – 3j. At the beginning, the population in level $|3\rangle$ increases, resulting in more population driven from $|3\rangle$ to $|2\rangle$ at longer time delay. Because the plasmon field is initially weak, it cannot drive the population from $|2\rangle$ to $|1\rangle$. Therefore the plasmon field will be amplified until it is strong enough to provide a significant feedback driving the $|2\rangle \rightarrow |1\rangle$ transition. Fig. 3e shows the sudden increase in the population inversion at the arrival time of the drive pulse.

However, when the delay $\tau_d > 0.5$ ps, the maximum number of plasmons decreases. This is because after 0.5 ps, the population inversion increases with the corresponding increase of the feedback leading to the early plasmon amplification before the arrival of the drive pulse (Fig. 3c). When the drive pulse arrives, its effect is influenced by the strong plasmon feedback which stimulates the $|2\rangle \rightarrow |1\rangle$ transition and decreases the population inversion (Fig. 3f). The corresponding plasmon and drive pulse profiles are shown in Fig. 4. Fig. 4b shows a small negative delay $\tau_0 \sim$-7 fs of the main plasmon pulse, and a small shoulder at earlier time due to the plasmon feedback. Similar behavior will takes place in the stable regime (Figs. 3d and 3g). The losses due to the plasmon feedback make it difficult to achieve the large population inversion which is necessary for giant plasmon pulse generation. The phase plane plots in Figs. 3h-3j show the presence of a large pulse followed by small oscillations. At longer delays, the giant plasmon pulse amplitude decreases and plasmon relaxation oscillations become more visible.

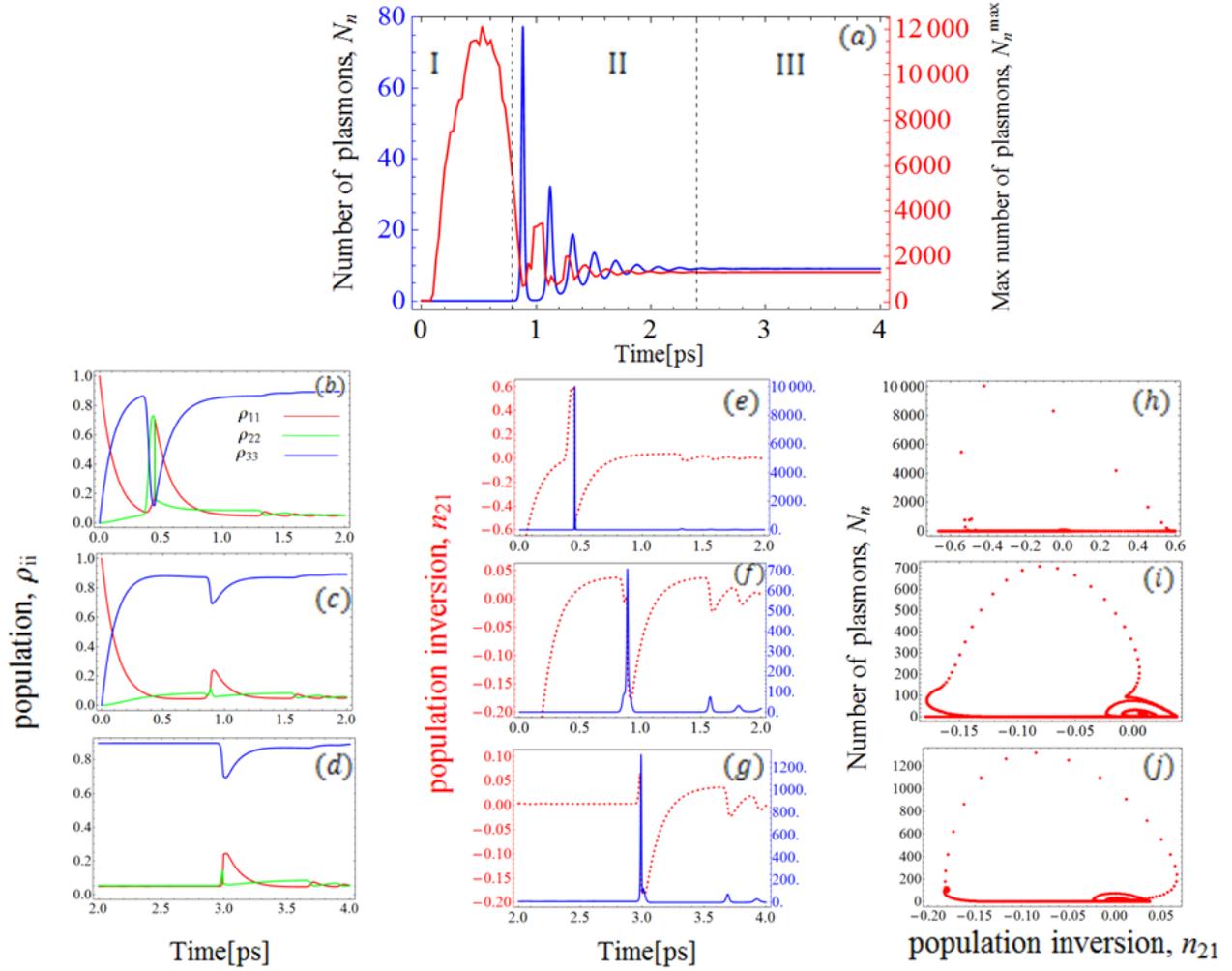

Figure 3. Control by the drive pulse delay. (a) Maximum number of plasmons (solid red) at different time delays of the Gaussian drive pulse with $\sigma_d=25$ fs and $\Omega_{a0}=24\times10^{12}$ s$^{-1}$ with respect to the switching of the pump compared with the number of plasmons without the drive (solid blue). Three regimes of plasmon pulse generation are separated by vertical dashed black lines: (I) the *first* regime of the largest plasmon pulse generation; (II) the second, *spiking* regime; and (III) the third, *stable* regime. Population evolutions $\rho_{ii}$, population inversion, $n_{21}$, plasmon pulse profiles, $N_n$, and phase plane plots for different time delays for the first $\tau_d=0.4$ ps ((b), (e) and (h)), the spiking $\tau_d=0.9$ ps ((c), (f) and (i)), and the stable $\tau_d=3$ ps ((d), (g) and (j)) regimes.

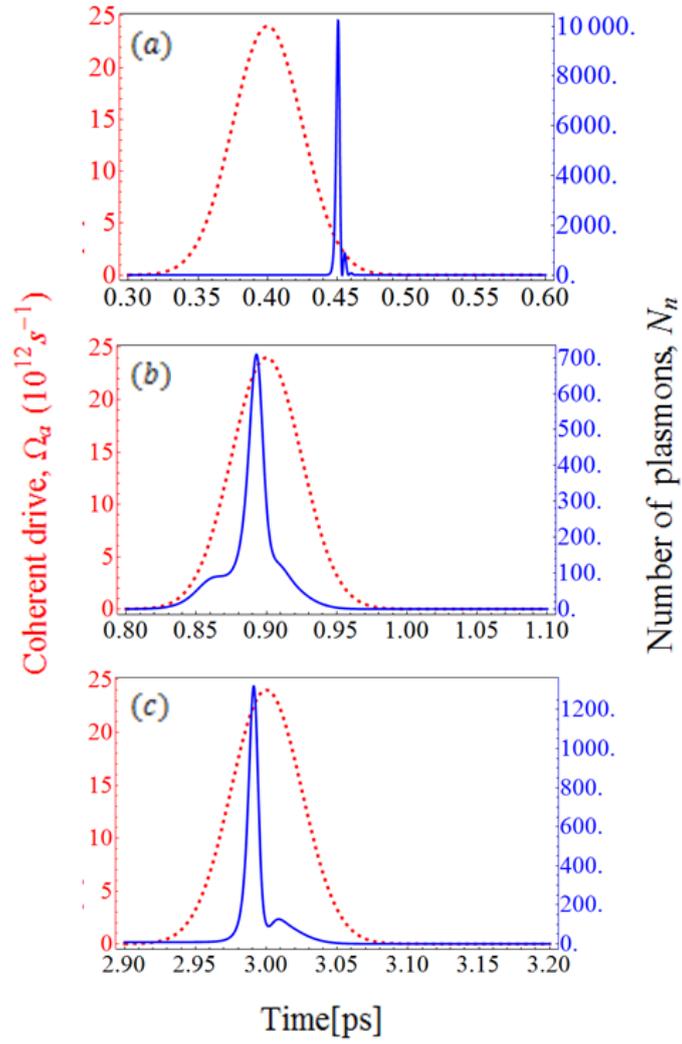

Figure 4. Temporal profiles of the number of plasmons, (solid blue) with a transient drive pulse (dashed red) arriving at different time delays for the first, $\sigma_d$=0.4 ps(a), the spiking, $\sigma_d$=0.9 ps(b), and the stable, $\sigma_d$=3 ps(c) regimes, corresponding to Fig. 3.

**Control by the drive pulse amplitude**

Next we investigate the quantum control of the plasmon pulse generation by varying the drive pulse amplitude. We consider the cases of the first and the stable regimes. Figs. 5a and 5b show the dependence of the maximum number of surface plasmons on the coherent drive pulse peak Rabi frequency $\Omega_{a0}$ for two values of the drive pulse duration $\sigma_d=25$ fs and $\sigma_d=75$ fs in the stable ($\tau_d=7.5$ ps) and the first ($\tau_d=0.625$ ps) regime, respectively. The number of plasmons increases nonlinearly with $\Omega_{a0}$. The nonlinearity depends on the pulse duration. The stronger drive transfers more population to level $|2\rangle$, creating a larger population inversion and generating stronger plasmon pulses.

The corresponding population inversion, plasmon pulse profiles, and phase plane plots for $\sigma_d=25$ fs for the stable and first regimes are shown in Figs. 5c-5j for $\Omega_{a0}=4\times10^{12}$ and $24\times10^{12}$ s$^{-1}$, respectively. In the stable regime, the drive pulse always increases the population inversion and generates stronger plasmon pulses without a delay between the drive pulse and the plasmon emission. However, in the first regime, when the drive pulse is weak with $\Omega_{a0}=4\times10^{12}$ s$^{-1}$ (Fig. 5g), the population inversion suddenly increases but the plasmon pulse is emitted at a later time. It takes time to develop a strong plasmon feedback to stimulate plasmon emission. However, a strong drive pulse with $\Omega_{a0}=24\times10^{12}$ s$^{-1}$ leads to a sudden generation of the plasmon pulse without a delay even in the first regime. The strong drive pulse efficiently generates the plasmon field which controls the generation of a strong plasmon pulse. The phase plane plots for the stable and first regimes reveal modified trajectories in phase space compared to the case without a drive. The plasmon pulse generation delay is clearly seen in Fig. 5i as a straight line segment at a large population inversion and small number of plasmons. The corresponding population evolutions are shown in Fig. 6. This behavior indicates the importance of the plasmon feedback for the control of relaxation oscillation spikes and formation of plasmon pulses.

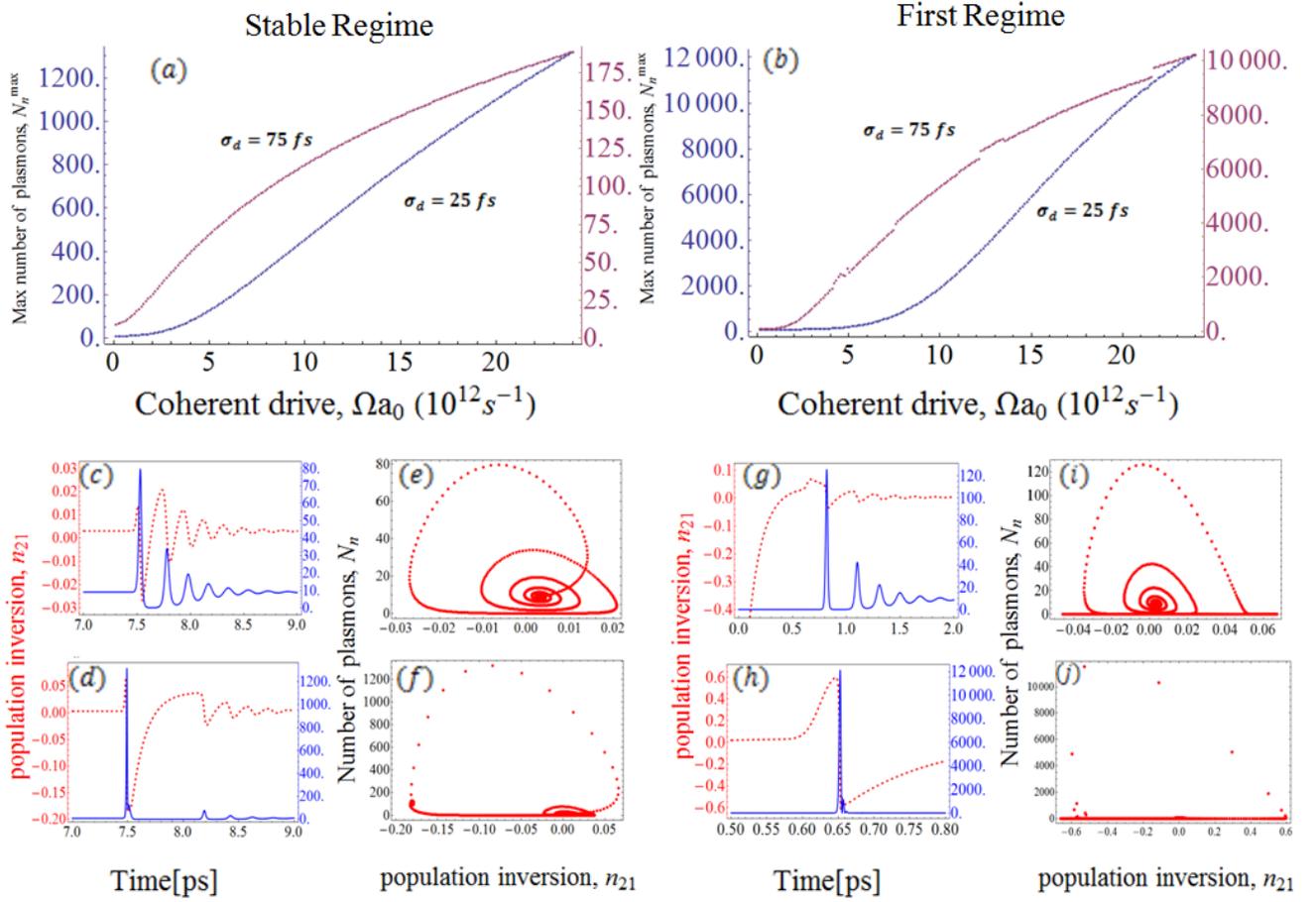

Figure 5. Control by the drive pulse amplitude. Maximum number of surface plasmons for the drive pulse duration $\sigma_d$=25 fs and $\sigma_d$=75 fs in the stable (a) and first (b) regime, respectively. The population inversion and plasmon pulse profiles for the stable (with $\Omega_{a0}$=4×10$^{12}$ s$^{-1}$ (c) and (with $\Omega_{a0}$=24×10$^{12}$ s$^{-1}$ (d)) and first (with $\Omega_{a0}$=4×10$^{12}$ s$^{-1}$ (g) and with $\Omega_{a0}$=24×10$^{12}$ s$^{-1}$ (h)) regimes (for the drive pulse duration $\sigma_d$=25 fs). The corresponding phase plane plots for the stable ((e) and (f)) and first ((i) and (j)) regimes.

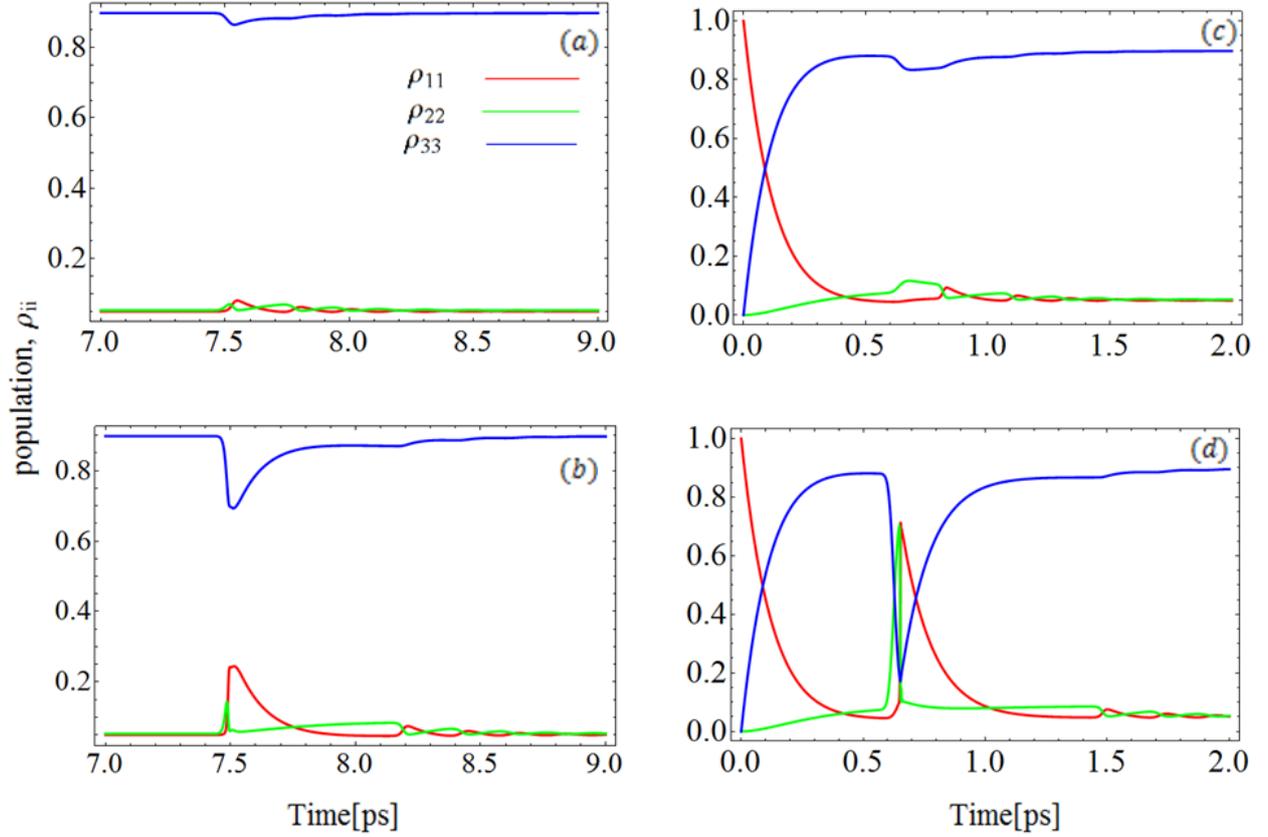

Figure 6. Population evolutions for the drive pulse duration $\sigma_d=25$ fs in the stable (with $\Omega_{a0}=4\times10^{12}$ s$^{-1}$ (a) and (with $\Omega_{a0}=24\times10^{12}$ s$^{-1}$ (b)) and first (with $\Omega_{a0}=4\times10^{12}$ s$^{-1}$ (c) and with $\Omega_{a0}=24\times10^{12}$ s$^{-1}$ (d)) regime corresponding to Fig. 5.

**Control by the drive pulse duration**

Finally we investigate the quantum control of the plasmon pulse generation by varying the drive pulse duration. Figs. 7a and 7b show the dependence of the maximum number of plasmons on the drive pulse duration (with $\Omega_{a0}=24\times10^{12}$ s$^{-1}$) in the stable ($\tau_d=7.5$ ps) and the first ($\tau_d=0.625$ ps) regime, respectively. Surprisingly, the number of plasmons increases dramatically with the decrease of the drive pulse duration (for the same drive pulse $\Omega_{a0}$) below 100 fs in the stable regime and below 400 fs in the first regime. This seems to be counterintuitive, meaning that using less drive energy leads to generating stronger plasmon pulses. Figs. 7a and 7b reveal three types of behavior. First, for the smallest drive pulse duration of less than ~ 5 fs the maximum number of plasmons increases with the drive pulse duration because longer drive pulse transfers more population from level |3⟩ to |2⟩ leading to larger population inversion and stronger plasmon pulses. However, in the second case, when the drive pulse duration is longer than ∼ 5 fs, only the first part of the drive pulse will contribute to the increase of the population inversion. After that the induced plasmon feedback will drive the population of |2⟩ to the ground state and will decrease both the population inversion and the maximum number of plasmons. In the third case, the drive pulse will make a small contribution to population inversion and will have a small effect on the coherent dynamics via the third term in Eq. (1). These effects can be explained as the competition between the two control mechanisms described above. The population-inversion mechanism (second term in Eq. (1)) dominates for short pulse durations and short delay times where population inversion is large. Longer pulse durations with small or no population inversion are dominated by the LWI mechanism (third term in Eq. (1)). The LWI mechanism generates long weak plasmon pulses whole temporal profile is similar to the profile of the drive pulse as can be seen in Figs. 7 and 8. Fig. 7 shows strong short plasmon pulses due to the population-inversion mechanism and long weak plasmon pulses due to the LWI mechanism. The phase plane diagrams show periodic spiraling behavior of plasmon relaxation oscillations for the positive population inversion due to the population-inversion mechanism, and chaotic irregular phase plane shapes in the negative population inversion due to the LWI mechanism.

In the first regime, the dependence of the maximum number of plasmons on the drive pulse duration is complicated, exhibiting irregular oscillations for $\sigma_d$ >400 fs. Fig. 7 also shows the population inversion, plasmon pulse profiles and phase plane plots for various drive pulse durations in the stable (c-h) and first (i-n) regimes for the pulse durations: $\sigma_d$=5 fs (c), 37.5 fs (d), 162.5 fs (e), 5 fs (i), 400 fs (j), and 25 ps (k). Peculiar behavior is observed in the phase plane plots for the long drive pulse durations in the first regime as the attractor point moves to the region of the more negative population inversion. This is due to the combined effect of the drive pulse duration and delay which leads to the comparable contributions of both control mechanisms. As a result, both the strong ultrashort plasmon pulses and weak long plasmon pulses are observed. The corresponding population evolutions are shown in Fig. 9.

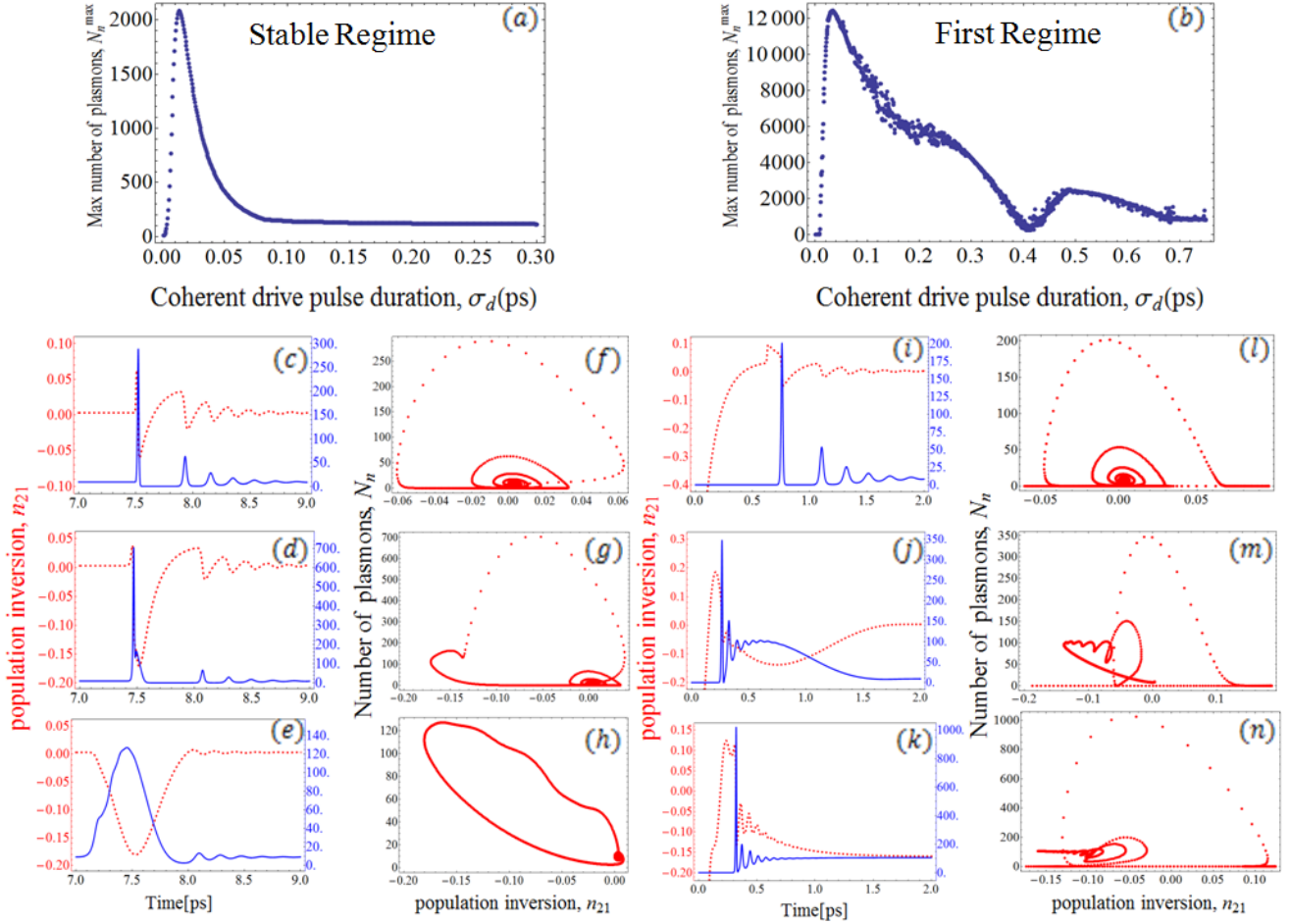

Figure 7. **Control by the drive pulse duration.** Maximum number of surface plasmons in the stable (a) and the first (b) regime, respectively. The population inversion and plasmon pulse profiles for various drive pulse durations in the stable regime: $\sigma_d$=5 fs (c), 37.5 fs (d), 162.5 fs (e); and in the first regime: $\sigma_d$=5 fs (i), 400 fs (j), and 25 ps (k). The corresponding phase plane plots for the stable ((f) – (h)) and first ((l) – (n)) regimes.

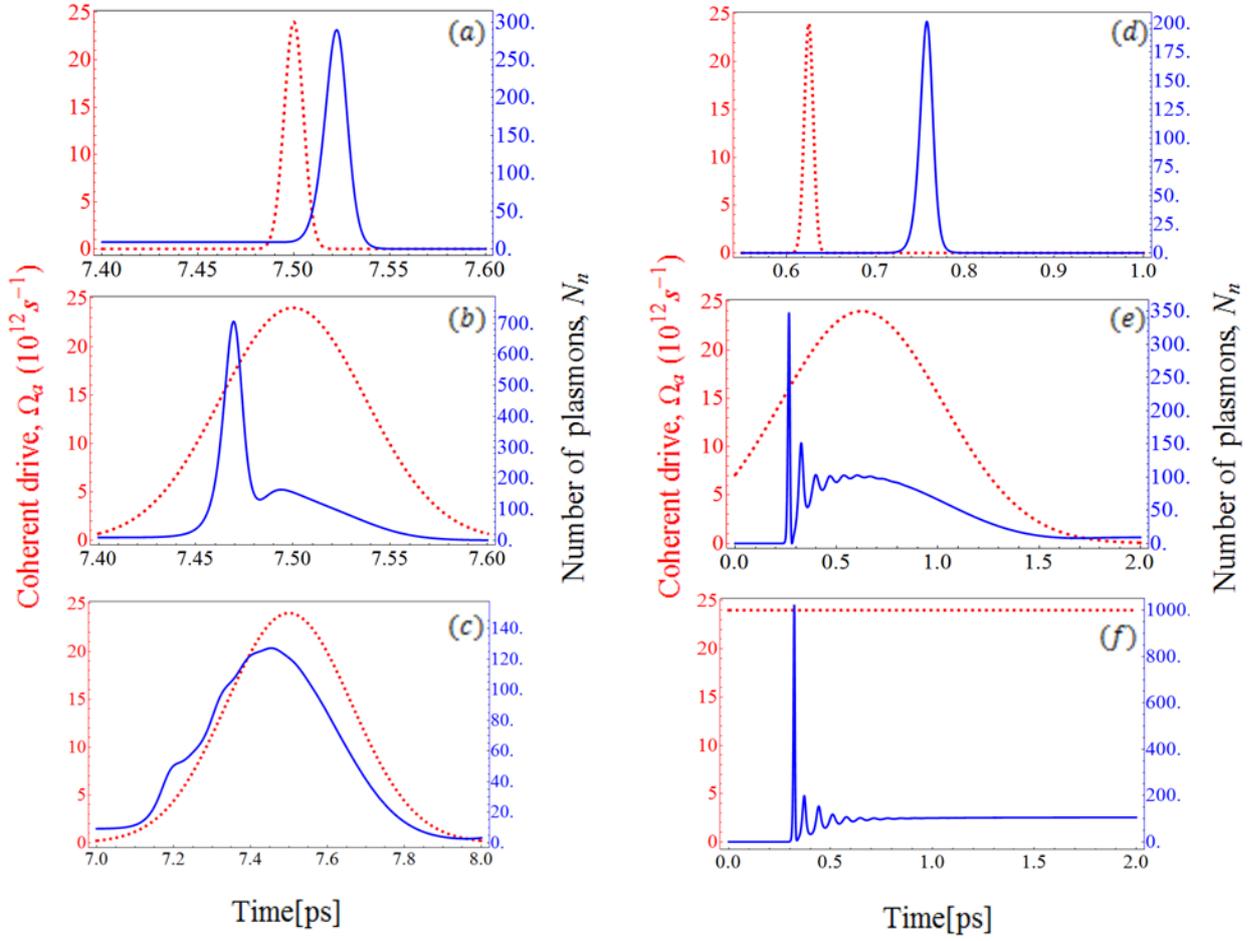

Figure 8. Temporal profiles of the number of plasmons, (solid blue) with a transient drive pulse (dashed red) for various drive pulse durations in the stable regime: $\sigma_d$=5 fs (a), 37.5 fs (b), 162.5 fs (c); and in the first regime: $\sigma_d$=5 fs (d), 400 fs (e), and 25 ps (f) corresponding to Fig. 7.

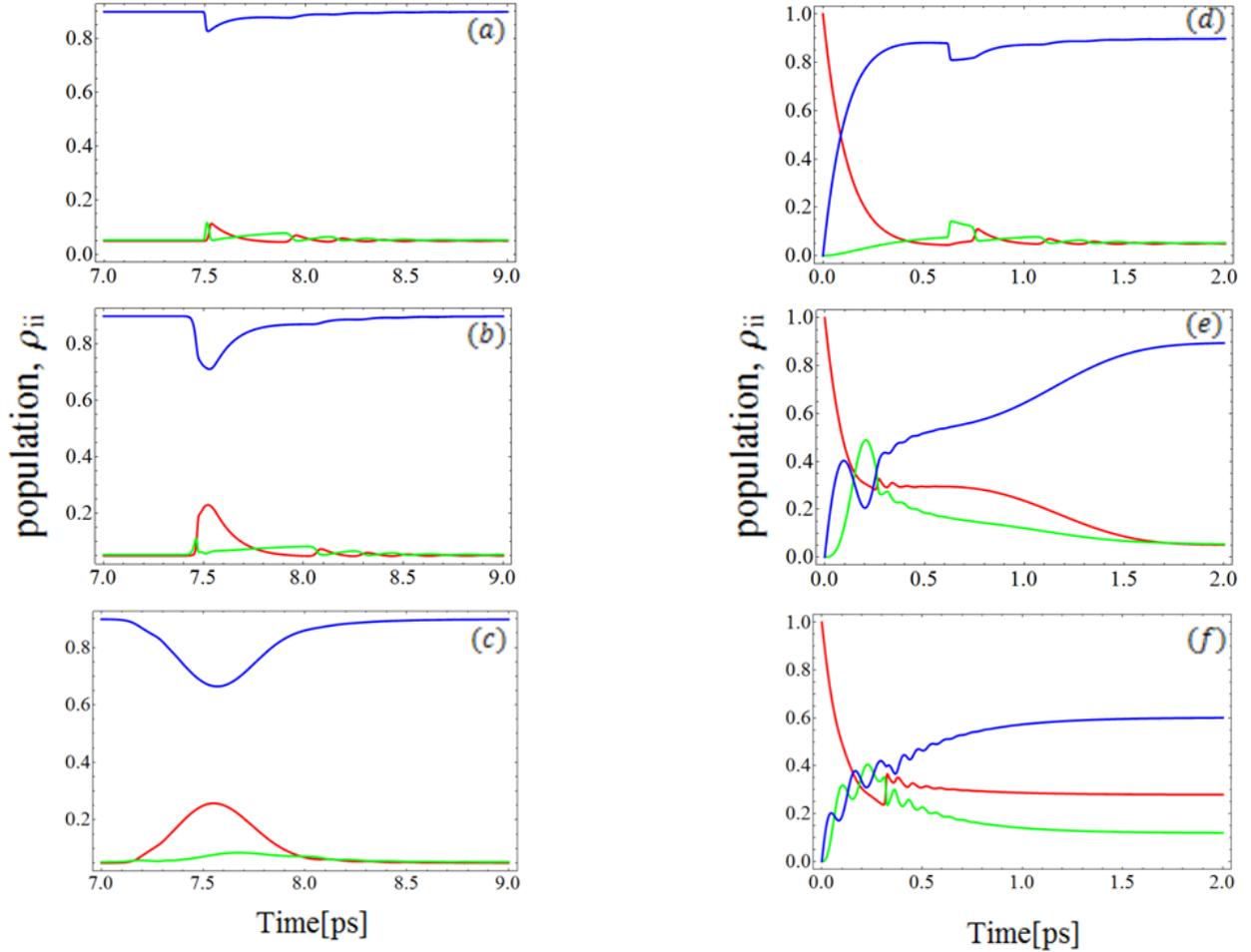

Figure 9. Population evolutions for various drive pulse durations in the stable regime: $\sigma_d$=5 fs (a), 37.5 fs (b), 162.5 fs (c); and in the first regime: $\sigma_d$=5 fs (d), 400 fs (e), and 25 ps (f) corresponding to Fig. 7.

## III. DISCUSSION

Figs. 3, 5 and 7 show the quantum control of plasmon population dynamics by varying the coherent drive pulse parameters. The most important control parameter for generation of ultrashort giant plasmon pulses is the time delay because it determines the optimal timing of switching the loss compensation by quantum coherence. If the drive pulse arrives too early, there is not enough population inversion and small amount of level $|3\rangle$ population. If the drive pulse arrives too late, the plasmon feedback is too strong. The drive pulse amplitude control parameter shows a nonlinearly increasing dependence. Higher drive pulse peak Rabi frequency leads to a larger number of plasmons. The drive pulse duration dependence is more complicated. It involves the pulse delay control parameter, especially in the first regime, which makes the interpretation challenging. The surprising

increase of the plasmon pulse peak intensity with the decrease of the drive pulse duration at the same peak Rabi frequency may be explained by the competition of two control mechanisms in Eq. (1). The population-inversion control mechanism can be used to modulate the gain medium losses on the ultrafast time scale and generate giant ultrashort plasmon pulses via quantum-coherent Q-switching. The drive pulse properties may be optimized to suppress the LWI control mechanism and to deposit more energy into the plasmon pulse. The resulting plasmon pulse duration is an order of magnitude shorter than the drive pulse duration. This can be used as a new method of short pulse generation, and may be combined in a cascade scheme, further reducing the pulse duration. The advantage of this approach to plasmon pulse generation compared to the direct excitation of plasmons via the surface plasmon resonance is that the direct excitation may lead to heating and melting of the nanostructure. This limits the performance to weak excitation intensities. Here we drive the system at a far-detuned transition which reduces the melting threshold.

Various practical implementation schemes of the proposed quantum-coherent ultrafast surface plasmon source are envisioned. One possibility is to use quantum dots coupled to plasmonic nanostructures as originally proposed by Stockman et al[32-35]. Specially designed quantum dots with two energy levels far-detuned from the plasmon resonance will be used to generate quantum coherence. The relaxation rates will have to be tuned to match our simulations. Another possibility is to use molecular aggregates coupled to plasmonic nanostructures which form so-called plexcitons[43]. A large number of molecules can form a layer of gain medium shown in Fig. 1a. Their energy levels and relaxation rates can be engineered using quantum chemistry and organic synthesis. For example, a special pair of chlorophylls in photosynthetic reaction centers can provide the two levels required to generate the quantum coherence. The reaction center was recently described as a quantum heat engine in analogy with lasers and solar cells, and its efficiency can in principle be increased by quantum coherence[44].

## IV. CONCLUSION

In summary, we have proposed a new, quantum approach to ultrashort pulse generation by transient quantum coherence. We applied it to generation of giant localized surface plasmon pulses in a silver nanosphere coupled to a three-level gain medium. We achieved an order of magnitude enhancement of the plasmon peak amplitude compared to the steady-state using a lower drive power, and two orders of magnitude enhancement compared to the case without a drive. We investigated the quantum control of plasmon dynamics by transient quantum coherence. We performed a few-parameter open-loop control by varying the drive pulse delay, amplitude and duration. Multi-parameter[45,46] and closed-loop[47] approaches may also be implemented in the future to further optimize the performance. Our approach can be applied to other ultrafast nanooptical generation and imaging techniques including surface plasmon polariton propagation[48], coherent surface-enhanced spectroscopy[49] and multidimensional nanoscopy[50].

## ACKNOWLEDGEMENTS

We gratefully acknowledge support of the National Science Foundation Grants PHY-1241032 (INSPIRE CREATIV), PHY-1068554, EEC-0540832 (MIRTHE ERC) and the Robert A. Welch Foundation (Awards A-1261). W.H. acknowledges the support of the Education Program for Talented Students of Xi'an Jiaotong University.